\documentclass[aps,twocolumn,showpacs]{revtex4}
\usepackage{graphics}
\usepackage{latexsym}
\usepackage{graphicx,epsfig}
\usepackage{psfrag}
\newcommand{\be}{\begin{equation}}
\newcommand{\ee}{\end{equation}}
\def\bq{\begin{eqnarray}}
\def\eq{\end{eqnarray}}

\begin{document}
\bibliographystyle{apsrev}
\def\half{{1\over 2}}
\def \D {\mbox{D}}
\def\curl {\mbox{curl}\,}
\def \ep {\varepsilon}
\def \lleq {\lower0.9ex\hbox{ $\buildrel < \over \sim$} ~}
\def \ggeq {\lower0.9ex\hbox{ $\buildrel > \over \sim$} ~}
\def\beq{\begin{equation}}
\def\eeq{\end{equation}}
\def\ber{\begin{eqnarray}}
\def\eer{\end{eqnarray}}
\def \apl {ApJ, }
\def \aps {ApJS, }
\def \pd {Phys. Rev. D, }
\def \prl {Phys. Rev. Lett., }
\def \pl {Phys. Lett., }
\def \np {Nucl. Phys., }
\def \l {\Lambda}  
\title{Phantom Field and the Fate of Universe }
\author{M. Sami}
\altaffiliation[On leave from:]{ Department of Physics, Jamia Millia, New Delhi-110025}
\email{sami@iucaa.ernet.in}
\affiliation{IUCAA, Post Bag 4, Ganeshkhind,\\
 Pune 411 007, India.}  
\author{Alexey Toporensky}
\affiliation{Sternberg Astronomical Institute, Universitetski Prospect, 13, Moscow 119899, Russia}
\email{lesha@sai.msu.ru}
\begin{abstract}
In this paper we analyze the cosmological dynamics of phantom field in a variety of potentials unbounded from above. 
We demonstrate that the nature of
future evolution generically depends upon the steepness of the phantom potential and discuss  the fate of Universe accordingly. 
\end{abstract}

\maketitle

\section{Introduction}
There is a growing evidence that Universe is undergoing accelerated expansion at present. The observations related to  supernova, CMB and galaxy clustering all together strongly point  towards the compelling possibility of late time accelerated expansion of Universe. The acceleration of Universe can be accounted for either by modifying the left hand side of Einstein equations or by supplementing the energy momentum tensor by an exotic matter with negative pressure, popularly known as dark energy. In past 
few years there have been tremendous efforts in modeling the dark energy. They include  scalar field models, some models of brane worlds and specific compactification schemes in string theory which have been
shown to  mimic the dark energy like behaviour. A wide variety of scalar field models have been conjectured for this purpose
including quintessence \cite{phiindustry}, K-essence\cite{Kes}, tachyonic scalar fields with the last one being originally motivated by string theoretical ideas. All these models of scalar field lead to the equation of state parameter $w$ greater than or 
equal to minus one. However, the recent observations do not seem to exclude values of this 
parameter less than minus one\footnote{ It is argued in Refs.\cite{jani,tp} that recent observations favor $w < -1$.}. It is therefore important to look for theoretical possibilities to describe dark energy
with $w <-1$ called phantom energy\cite{ph0,ph1,ph3,ph4,ph5,ph6,ph7,ph8,ph9,ph10,hann,t1,t2,t3,ph12,ph13,ph14,ph15,ph16,ph17,ph18,
ph19,ph20,phs,
ph21,ph22,ph23,ph24,ph25,ph26,ph27,ph28,ph29,sashak,piao,jani,tp}. Specific models in brane world or non-minimally coupled scalar
fields may lead to phantom energy \cite{phsh,nm}. In our opinion, the simplest alternative is provided by a phantom
field, scalar field with negative kinetic energy. Such a field can be motivated from
S-brane constructs in string theory\cite{sb,sb1,sb2,sb3,sb4,IN1,IN2,IN3}. Historically, phantom fields were first introduced in
Hoyle's version of the Steady State Theory. In adherence to the Perfect Cosmological Principle, a creation field (C-field) was for the first time introduced \cite{h} to reconcile with homogeneous density by creation of new matter in the voids caused by the expansion of Universe. 
It was further refined and reformulated in the Hoyle and Narlikar theory of gravitation \cite{hn}(see also Ref.\cite{pn} on the
similar theme).
Though the quantum theory of phantom fields is problematic\cite{ph30}, it is nevertheless interesting to examine the cosmological
consequences of these fields at the classical level.\par                                                              
 Models with constant $w<-1$ lead to unwanted future
singularity called {\it big rip}\cite{staro}. This singularity is characterized by the divergence of  scale factor after a finite
interval of time. Infact there is no profound reason to assume that the equation of state parameter $w$
would remain constant in future if it has been changing all the way since the end of inflation to date. Keeping
this in mind, specific scalar field models were proposed to avoid the cosmic doomsday\cite{t2,phs}. It requires a special
class of phantom field potentials with local maximum.   
In this paper we examine the nature of future evolution of Universe with a phantom field evolving in a more general class 
of potentials. The future course of evolution is shown to critically dependent on the steepness of the underlying potential. 
\section{Cosmological dynamics in the presence of Phantom Field}
The Lagrangian of the phantom field minimally coupled to gravity and matter
 sources is given by\cite{ph15,phs}
\begin{equation}
{\cal{L}} =(16 \pi G)^{-1} R+{1 \over 2} g^{\mu \nu} \partial \phi_{\mu} \partial \phi_{\nu}-V(\phi)+{\cal{L}}_{source}
\label{lagrang}
\end{equation}
where ${\cal{L}}_{source}$ is the remaining source term (matter, radiation)
 and $V(\phi)$ is the phantom potential. The kinetic energy term of the phantom
  field in (\ref{lagrang})  enters with the opposite sign in contrast to
   the ordinary scalar field (we employ the
metric signature, -,+,+,+). In a spatially flat FRW cosmology, 
the stress tensor that follows from (\ref{lagrang}) 
acquires the diagonal form $T^{\alpha}_{\beta}=diag\left(-\rho,p,p,p\right)$
 where the pressure and energy density of field $\phi$ are given by
\begin{equation}
\rho_{\phi}=-{\dot{\phi}^2 \over 2}+V(\phi),~~~~~p_{\phi}=-{\dot{\phi}^2 \over 2}-V(\phi).
\label{ke}
\end{equation} 
The corresponding equation of state parameter is now given by
\begin{equation}
w \equiv {p_{\phi} \over \rho_{\phi}} ={{{\dot{\phi}^2 \over 2}+V(\phi)} \over { {\dot{\phi}^2 \over 2}-V(\phi) }}.
\end{equation}
For $\rho_{\phi} >0$, $w <-1.$

 The equations of motion which follow from (\ref{lagrang}) are 
 \begin{equation}
 \dot H=-\frac{1}{2 M_p^2}(\rho_b+p_b-\dot \phi^2)
 \label{doth}
 \end{equation}
\begin{equation} 
H^2=\frac{1}{3 M_p^2}(\rho_b+\rho_{\phi})
\label{hubble}
\end {equation}  
\begin{equation} 
\ddot \phi+3H\dot \phi = V'(\phi) 
\label{fieldeq}
\end{equation}
where the background energy density due to matter and radiation is given by
\begin{equation}
\rho_b={\rho^i_R \over a^4}+{\rho^i_m \over a^3}
\end{equation}    
 
\begin{figure}
\resizebox{3.0in}{!}{\includegraphics{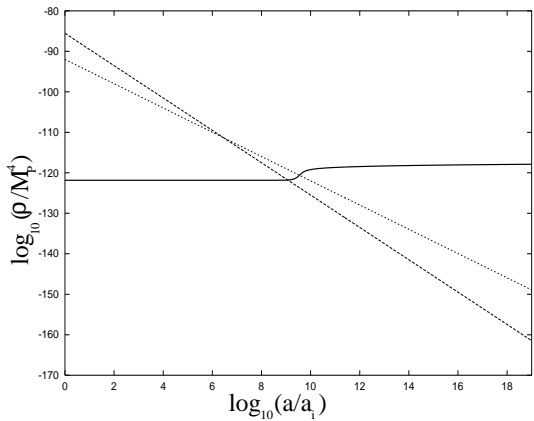}}
\caption{Evolution of  energy density is plotted against the scale factor. Solid line corresponds to the
phantom field energy density  
in case 
of the model described by the potential: $V(\phi)=m^2 \phi^2 $ with $m \simeq 10^{-60} M_p $. The dashed and dotted lines  correspond to 
energy density of radiation and matter.
Initially, the phantom field mimics the cosmological constant like behaviour and its energy density 
is extremely subdominant to the background (the system is numerically evolved starting from the radiation domination
era with $\rho_r =1 MeV^4$) and remains to be so for most of the
period of evolution. The phantom field $\phi$ continues in the state with $w=-1$ till the moment $\rho_{\phi}$ approaches $\rho_b$. The background ceases now to play the leading role (becomes subdominant) and the phantom field
takes over and starts climbing up the potential fast. 
At late times, the field energy density catches up with the background, overtakes it and starts growing ($w <-1)$
and drives
the current (the value of the scale factor $ a \simeq 4 \times 10^{9}$ corresponds to the present epoch) accelerated expansion of the Universe with $\Omega_{\phi} \simeq 0.7$ and $\Omega_m \simeq 0.3$. Initial value of the field was tuned to get the
present values of $\Omega$. The field
then enters into the slow climbing regime allowing the slowing down of growth of $\rho_{\phi}$ and making $w \to -1$
asymptotically which is an attractor in this case.
}
\label{PDEN}
\end{figure}

As in the previous work \cite{t2,phs} we start with the phantom field energy density
 being dominated by the density of the ordinary matter. 
In this situation, an initial kinetic
term of the field  rapidly decreases due to the friction term $3 H \dot{\phi}$
in (\ref{fieldeq}) and as a result the phantom field "freezes" at some point waiting for the moment
its energy density becomes comparable with the matter (see figure \ref{PDEN}). Subsequently,
the field dynamics "switches on" and the nature of future evolution  then depends on the
shape of the phantom field potential $V(\phi)$. \par 
The case of potentials with a local maximum was already described in 
\cite {t2,phs}. The phantom field asymptotically reaches the maximum with a
possible oscillating transient stage. The future asymptotic regime is the De-Sitter
one with $\omega \to -1$ and $\Lambda = V_{max}/M_p^2$.

 Potentials other than these can lead  to a variety of future evolutionary regimes.
We begin with a simplest potential of a massive phantom field 
\begin{equation}
V(\phi)=m^2 \phi^2 
\end{equation}
When the phantom energy begins to dominate and $\dot \phi$ is small,
 we have, see equation (\ref{hubble}) and (\ref{fieldeq}) 
\begin{equation}
H \simeq {{m\phi} \over {\sqrt{3 M_p^2}}},~~~~~~   \dot \phi \simeq {{2mM_p} \over \sqrt{3}}
\end{equation} 
Once such a regime is established its stability is guaranteed (see figure 2). Indeed, the
 ratio $\dot{\phi}^2/2V$ of the phantom field is inversely
 proportional to $\phi^2$ and tends to zero making  
 the kinetic term subdominant forever. This regime is analogous to the
  slow-roll regime for a normal scalar field and can be named as "slow-climb".
The effective equation of state tends to the De-Sitter one (see figure 3) with a slow growing 
energy density(figure 1). As a result, the final state of such Universe is formally
different from both the De-Sitter and the "big rip" outcome - it would take 
 infinite time to reach an infinite energy density in this case. However,
such Universe will steadily reach the Planck density for the finite time and 
the classical physics breaks down here.
                
It can easily be checked that the condition $\dot{\phi}^2/2V  \to 0$
is satisfied for any potential, which asymptotically reduces to power-law.
Indeed, if the kinetic term is subdominant, equation (\ref{fieldeq})  gives
\begin{equation}
\dot \phi=\frac{V'(\phi)}{3H}
\label{phidot1}
\end {equation}
In the regime of phantom field dominance we have 
\begin{equation}
H^2=\frac{V(\phi)}{3 M_p^2}
\end{equation}

\begin{equation}
\left(\frac{\dot \phi^2}{2}\right) {1 \over V(\phi)}={M_p^2 \over 6} \left(\frac{V'^2}{ V^2}\right)
\label{phidot}
\end{equation}
It is clear from (\ref{phidot}) that $\dot{\phi}^2/2V$  is proportional to $\phi^{-2}$ for any power-law potential
$V(\phi) \sim \phi^{\alpha}$.

\begin{figure}
\resizebox{3.0in}{!}{\includegraphics{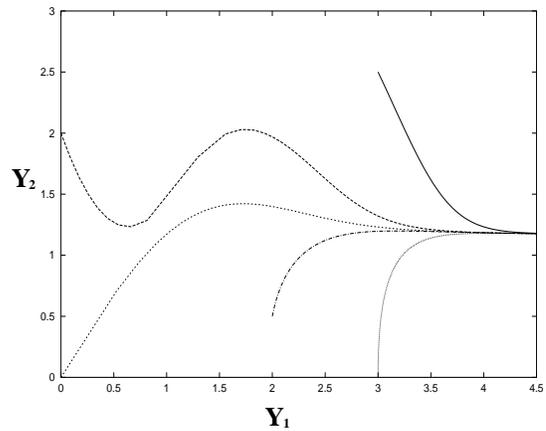}}
\caption{Display of the phase portrait (plot of $Y_2 \equiv \dot{\phi}/M_p^2$ versus $Y_1 \equiv \phi/M_p$)
for the phantom field in case of the quadratic potential $V(\phi)=m^2 \phi^2$. The figure shows that the trajectories
starting anywhere in the phase space move towards a configuration({\it slow climb} regime) with 
$\dot{\phi} \to {\it const}$ asymptotically(the convenient choice of mass parameter in the potential used here corresponds
to the value of the {\it const} equal to $2/\sqrt{3}$). This picture is drawn
in absence of the background energy density and allows to probe a wider class of initial conditions; the asymptotic regime is independent of the background.
}
\label{PEQSTAT}
\end{figure}

\begin{figure}
\resizebox{3.0in}{!}{\includegraphics{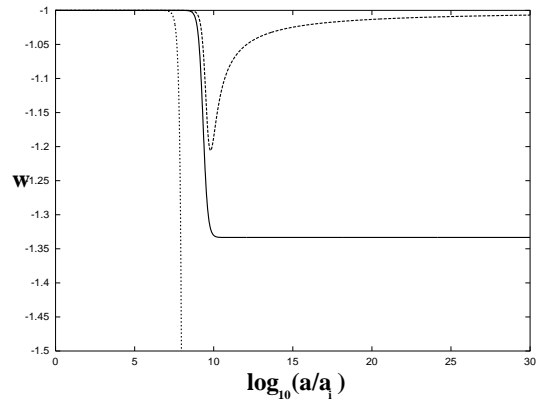}}
\caption{Equation of state parameter $w$ for the phantom field is plotted against the scale factor. The solid line
corresponds to the exponential potential $V(\phi) \sim e^{\lambda \phi/M_p}$, the dotted line to $V(\phi) \sim e^{\lambda \phi^2/M_p^2}$ and the dashed line to $V(\phi) \sim \phi^2$. The exponential potential exhibits a critical behaviour in 
which case $w$ fast evolves to a constant value less than minus one (numerical value depends upon $\lambda$ which was taken to be equal to one for convenience). For potential steeper than the exponential $w$ fast evolves towards larger and larger
negative values (dotted line) leading to a singularity
at some finite value of the scale factor. In case of  a less steeper potential
than the exponential, the equation of state parameter $w$ first decreases fast and then subsequently turns back towards
minus one and approaches it asymptotically (dashed line). 
}
\label{PPHASE}
\end{figure}  
However, the formally "slow climb" regime for steeper power-law potentials becomes
the regime of a fast growth of the phantom field. The potential $V = \lambda \phi^4$
leads to
\begin{equation} 
H \simeq {{\sqrt{\lambda} \phi^2} \over {\sqrt{3 M_p^2}}} ,~~~~~~~ V' =4\lambda \phi^3
\end{equation}
corresponding to 
\begin{equation}
\dot \phi \simeq 4 M_p {{\sqrt{\lambda \over 3}}} \phi
\end{equation} 
which results in exponential growth of $\phi$.
A bigger power index $\alpha$ leads to a "classical big rip" when
a formally infinite energy density is reached during a finite time.
Indeed, the general "slow-climb" equation for the phantom field with
a power-law potential
\begin{equation}
\dot \phi \simeq \phi^{\alpha/2-1} 
\end{equation}
in case of $\alpha > 4$ leads to
\begin{equation} 
\phi \simeq
(t_0-t)^{2/(4-\alpha)}
\end{equation}                     
For completeness we note that asymptotically flat potentials,
like $ V(\phi)=A(1-\exp{(-c \phi^2)})$ also lead to a "slow climb" but with 
asymptotically De-Sitter like behavior ($\Lambda = V(\infty)/M_p^2$).\par
The approximation $\dot{\phi}^2/2V \to 0$ breaks down for exponential and steeper potentials. 
In case of  exponential potentials\cite{ph16}, Hao and Li found an attractor solution with
  $\omega$ tending
to a constant value less then $-1$ making the "big rip" inevitable.

When we turn to potentials steeper than an exponential,
like $V(\phi)=\exp{(c \phi^2)}$ or potentials with an infinite
potential wall,  we get another
type of a future singularity. It is evident from equation (\ref{phidot}) that the
kinetic term grows faster than the potential one, and at a particular epoch
 becomes important.
 However, $\rho_{\phi} 
=V(\phi)-\dot \phi^2/2$ always grows as
the equation (\ref{doth}) in the epoch of phantom dominance
prevents $H$ and correspondingly  $\rho_{\phi}$
(see equation (\ref{hubble})) from decreasing. As a result, both potential and kinetic
terms increase rapidly with less rapid increase of $\rho_{\phi}$. Our numerical studies
indicate that the ratio of kinetic and potential terms tends to $-1$
and the state equation parameter $\omega$ blows up to $-\infty$ at some
finite value of the scale factor $a$ (figure 3). Potential and kinetic terms of the phantom
field also become infinite at this moment.
This type of cosmological singularity,
when equations of motion  become singular at a finite value of scale factor
are already known in some models of brane worlds\cite{shsa}, 
tachyonic field\cite{sashak} and
Gauss-Bonnet cosmology \cite{lesha}. However, we should note that
all these scenarios lead to divergence in the second derivatives of metrics.
The matter content of the Universe and Hubble parameter remain finite in these models. In case 
of the phantom field, the energy density itself blows up at some finite value of $a$.

\section{Conclusions}                                
We have studied the possible future regimes of Universe
 with a phantom field. The opposite sign in the kinetic term
  pushes the field towards maximum of its potential. 
   This feature leads to a cosmological
    singularity in the future for phantom
   field potentials unbounded from above. The nature of this singularity crucially depends 
     upon steepness of the potential. The analysis presented above allows us to classify the future
      cosmological behavior in the following types:
       \begin{itemize}
        \item For asymptotically power-law potentials $V(\phi) \sim \phi^{
	 \alpha}$ with $\alpha \le 4$ we obtain $\omega \rightarrow -1$ and $\rho_{\phi}
	  \to \infty$ for $t \to \infty$.
	   \item For $\alpha > 4$ though $\omega \to -1$, but the infinite energy
	    density is reached during a finite time, referred to as {\it big rip}.
	     \item For exponential potentials, $\omega$ tends to a constant value
	      less than $-1$ again leading to a {\it big rip} singularity.
	       \item Potentials steeper than exponential lead to a singularity with $w \to -\infty$ for a finite value of scale factor.
	     
		 \end{itemize}               

\begin{acknowledgements}
We thank N. Dadhich, Parampreet Singh and V. Sahni
for useful discussions. We also thank A. Starobinsky, Yun-Song Piao and  Ishwaree P. Neupane for comments. MS is thankful to T. Padmanabhan for helpful discussions. 
AT acknowledges support from IUCAA's
`Program for enhanced interaction with the Africa-Asia-Pacific Region'.
AT is grateful to IUCAA for hospitality where this work was done.

\end{acknowledgements}

\end{document}